\documentclass[runningheads]{llncs}
\usepackage{graphicx}
\usepackage{hyperref}
\usepackage{caption}
\usepackage{microtype}
\usepackage{booktabs}
\usepackage{multirow} 
\usepackage{marvosym}
\usepackage[numbers,sort&compress]{natbib}
\usepackage[section]{placeins}
\usepackage{amssymb}
\usepackage{pifont}
\usepackage{svg}
\usepackage{pdfpages}

\begin{document}
\title{The Importance of Distrust in AI 
\thanks{We gratefully acknowledge funding from 
the Deutsche Forschungsgemeinschaft (DFG, German Research
Foundation) for grant TRR 318/1 2021 – 438445824.\\
This preprint has not undergone peer review or any post-submission improvements or corrections.
The version of records of this contribution is published in Explainable Artificial Intelligence First World Conference, xAI 2023, Lisbon, Portugal, July 26–28, 2023, Proceedings, Part III (CCIS, volume 1903) and is available at https://doi.org/10.1007/978-3-031-44070-0  .}} 
% \thanks{Acknowledgement REDACTED}}
%
\titlerunning{Importance of Distrust in AI}
\author{Tobias M. Peters\inst{1}\orcidID{0009-0008-5193-6243}$^{\dag}$ \and Roel W. Visser\inst{2}\orcidID{0009-0006-3067-5545}$^{\dag}$}
% \author{REDACTED AUTHORS}
%
\authorrunning{T.M. Peters and R.W. Visser}
\institute{Cognitive Psychology\\Paderborn University, 33098 Paderborn, Germany\\
\email{tobias.peters@uni-paderborn.de}\\
\and
CITEC - Cognitive Interaction Technology\\Bielefeld University, 33619 Bielefeld, Germany\\
\email{rvisser@techfak.uni-bielefeld.de}}
\maketitle              % typeset the header of the contribution
\def\thefootnote{\dag}\footnotetext{These authors contributed equally to this work and share first authorship.}\def\thefootnote{\arabic{footnote}}
\begin{abstract}

In recent years the use of Artificial Intelligence (AI) has become increasingly prevalent in a growing number of fields. As AI systems are being adopted in more high-stakes areas such as medicine and finance, ensuring that they are trustworthy is of increasing importance. 
A concern that is prominently addressed by the development and application of explainability methods, which are purported to increase trust from its users and wider society. While an increase in trust may be desirable, an analysis of literature from different research fields shows that an exclusive focus on increasing trust may not be warranted. Something which is well exemplified by the recent development in AI chatbots, which while highly coherent tend to make up facts. In this contribution, we investigate the concepts of trust, trustworthiness, and user reliance.

In order to foster appropriate reliance on AI we need to prevent both disuse of these systems as well as overtrust. From our analysis of research on interpersonal trust, trust in automation, and trust in (X)AI, we identify the potential merit of the distinction between trust and distrust (in AI). We propose that alongside trust a healthy amount of distrust is of additional value for mitigating disuse and overtrust. We argue that by considering and evaluating both trust and distrust, we can ensure that users can rely appropriately on trustworthy AI, which can both be useful as well as fallible.

\end{abstract}
%
%
%
%
%
%%%%%%%%%%%%%%%%%%%%%%%%%%%%%%%%%%%%%%%%%%%%%%%%%%%%
%%%%%%%%%%         Introduction     %%%%%%%%%%%%%%%%
\section{Introduction} \label{sec:intro}

Intelligent systems and decision making supported by Artificial Intelligence (AI) are becoming ever more present and relevant within our everyday lives. 
Especially their use in high-stakes applications like medical diagnosis, credit scoring, and parole and bail decisions have led to concerns about the AI models \cite{rudin2019stop}. 
This includes concerns about the AI's transparency, interpretability, and fairness \cite{Guidotti2019, Arrieta2020, Mohseni2021}. These objectives are acknowledged and further enforced in legislation by the EU's General Data Protection Regulation (GDPR, Art. 15), in which citizens are granted the right to be provided with meaningful information about the logic involved in automated decision making.

While contemporary AI methods are becoming increasingly accurate and sophisticated, they often remain opaque and may, and most likely will, produce errors.
For instance, recent development in generative AI chatbots have highlighted that there remains a risk in relying on AI. 
While the current transformer-based large language models (LLMs) are very good at generating highly convincing and coherent texts, they are known to make up facts and can be inaccurate to the extent of fabricating entire quotes and references \cite{Ji2023}. 
While it may be appropriate to use such models in certain low-stakes applications, their inherent fallibility is more problematic in, say, a medical setting.

These developments and different concerns have lead to increased research interest into making AI systems more trustworthy and reliable.
One prominent way to address these growing concerns and new objectives is for modern (blackbox) AI methods to be able to explain their outputs \cite{Arrieta2020}, leading to a surge in the development of explainable AI (XAI) over the last years for a host of different applications, domains, and data types \cite{Guidotti2019, Arrieta2020, Samek2021}. 
Additionally, a number of different guidelines have been set out to ensure the trustworthiness of AI (for an overview see \cite{Thiebes.2021}),
the main objective being that ensuring the trustworthiness of AI should help increase user trust. 
Likewise, in the literature explainability is often explicitly considered as a means to increase user trust \cite{Kastner.2021}. 

In this contribution, we take a closer look at the connection between user trust and trustworthiness and explainability and its limitations with respect to insights from psychology. Currently, such insights are often incorporated only superficially or founded more on common-sense reasoning on trust. 
As even the best-performing models are prone to errors, we argue not to focus exclusively on increasing trust, but rather on establishing an appropriate level of reliance from the user on the AI with a \textbf{healthy amount of critical reflection, or distrust}, along with a \textbf{sufficient level of trust}. 
In the following, we investigate why not only increasing trust but also taking into account the importance of distrust is relevant for appropriate reliance on AI, preventing both the disuse as well as overtrust of such systems. 
For this, in the next sections we give an overview of literature related to automation, AI, human-computer interaction, and Psychology, in order to establish the relation between and importance of trust, distrust, reliance, and trustworthiness of AI. Some things which are of primary concern in the employment of XAI. 

\section{Trusting an AI?}  

Trustworthy AI is defined according to a number of design objectives that AI should conform to in order to be trustworthy to its users and to wider society \cite{Arrieta2020}.
An example of these are those formulated in the Ethics Guidelines for Trustworthy AI \cite{hlegAI}. 
The exact definition of which design objectives should be taken into account and which concerns should be addressed depends, for example, on who the concerning party is (e.g. the European commission in place of wider society) or to whom it is addressed to (e.g. end users vs machine learning engineers) \cite{Arrieta2020, Mohseni2021}. 
Of these objectives transparency and interpretability are the most important ones in the context of trust and XAI \cite{Arrieta2020, Guidotti2019, Mohseni2021}. 

\subsection{Explainability and Trust}
\label{sec:explainability-trust}
The development of explainable AI methods is currently one of the most prominent ways of working on means for addressing these concerns and fulfilling such regulations \cite{Toreini2020, langer2021}.
It is focused on developing ways to make AI both (more) interpretable and transparent, thereby ensuring that both its users and wider society can trust that an AI will work in a way that is intended, expected, and desirable \cite{Arrieta2020}. 
A multitude of XAI studies implicitly or explicitly assume explainability to facilitate trust \cite{Kastner.2021, ferrario2022}. 
In their summary of current XAI studies concerned with user trust, \citet{Kastner.2021} call this the explainability-trust hypothesis.
In contrast, results of empirical investigations into the observed relation between explanations and trust are summarized by \citet{Kastner.2021} as mixed and inconclusive. These results range from positive relations to no effect up to negative relations, which calls the validity of the explainability-trust hypothesis into question.
Similarly, \citet{ferrario2022} have serious doubts about the usefulness of explainability methods in fostering trust in the case of medical AI applications, and note that the relation between the explainability of AI and trust are far from being clarified.

One potential reason, which \citet{Kastner.2021} also entertain, of why explanations could fail to foster trust is that explanations can actually reveal problems of the system that may have otherwise gone unnoticed, which could lead a user not to trust the AI. 
To reveal problems of an AI is a function of explanations that is also targeted in the paper on the LIME algorithm, one of the first popular XAI methods which has been used for deep models \cite{Ribeiro2016}. 
According to \citet{Ribeiro2016} explanations are not only helpful for deciding when one should trust a prediction, but also beneficial in identifying when not to trust one. Thereby, they differentiate between the explanation's utility for trusting and not trusting, demonstrating the latter in an example where an explanation reveals a wrong causal relation underlying an automated decision. Yet, when generally discussing the benefit of explanations, \citet{Ribeiro2016} argue that ``[...] explaining predictions is an important aspect in getting humans to trust and use machine learning effectively, if the explanations are faithful and intelligible". With the conditional part of the sentence they acknowledge the possibility of explanations to indicate erroneous predictions, but still mainly focus on convincing a human to trust. This sets the focus on the utility of explanations to identify correct predictions, while the utility of explanation to identify wrong predictions falls short. 

This pattern can be found across much XAI literature that discusses user trust. 
When authors speak in broad terms, they connect explanations to the facilitation of trust, which represents the explainability-trust hypothesis as discussed above. Thereby, the explanation's utility to indicate correct predictions is discussed because only in this case a facilitation of trust is desired. 
However, when authors describe the actual utilities of explanation methods, another utility of explanations is also identified, e.g., not trusting predictions \cite{Ribeiro2016}, critical reflection \cite{Ehsan.2020}, or enabling distrust \cite{Jacovi.2021}. 
To clarify the aim of these different utilities, some XAI research employs the terms disuse and overtrust \cite{Jacovi.2021, Mohseni2021}.
Before we discuss both utilities of explanations and their connection to disuse and overtrust in more detail in Section \ref{sec:reliance}, it is first necessary to disentangle the terms trust and trustworthiness (Section \ref{sec:trustworthinessTrust}).

This necessity partially stems from the fact that trust is a typical objective of XAI, but often in papers concerning the development of XAI methods no proper definitions of trust are given \cite{Ribeiro2016, Gunning.2019, Bansal.2021}. E.g., \citet{Ribeiro2016} only separate trusting a prediction and trusting a model without explicating what is meant by trusting. Comparably, \citet{ferrario2022} observed that the dynamics between trust and explainability are far from being clarified, which they primarily attribute to the lack of precise definitions of the two. 
A summary of important terms in the AI context, which we discuss in the following, is provided in Table \ref{tab:terms}. 

Setting trust as a goal of explainability without providing information on what trust encompasses can lead to different problems. Firstly, in empirical investigations, cooperation or confidence might be easily mistaken for trust (see Section \ref{sec:Trust}). Secondly, without drawing from previous work on trust and their definitions, trust in the (X)AI context runs the risk of falling back into the state of a conceptual confusion regarding the meaning of trust \cite{Lewis.1985} that earlier work on trust aimed to overcome. 
Furthermore, basing trust in (X)AI research on the already established definitions of trust, the desired outcome, i.e. trust, becomes more standardized. This can improve the comparability between different studies, which then can allow researchers to make more general assumption about improvements of AI and their effect on trust. In addition, the process of evaluating potential effects of explainability on trust may profit as well. 

%%% table of terms
\begin{table}[h]
    \centering
    \begin{tabular}{|l|p{7.5cm}|}
        \hline
        Term               & Definition                                                                                   \\
        \hline
        Trustworthy AI     & Descriptive term for a desired form of AI.\\[2ex]  
        Reliance           & A human decision or action that considers the decision or recommendation of an AI.             \\[2ex]
        Trustworthiness    & Property of an AI, which leads an interactor to trust the system.                            \newline
                             Property of an AI, which justifies to place trust in it \cite{Toreini2020}. \\[2ex]
        Trust              & The willingness of a person to rely on AI in a situation that involves risk and uncertainty. \newline
                             ``Trust is an attitude a stakeholder holds towards a system." \cite{Kastner.2021}             \\[2ex]
        Appropriate Trust  & ``User’s ability to know when to, and when not to, trust the system’s
                             recommendations and decisions." \cite{Gunning.2019} 
                                    \\
        \hline
    \end{tabular}
    \vspace{0.5cm}
    \caption{Definitions of important terms in the AI context.}
    \label{tab:terms}
\end{table}

%%%%%%%%%%                          %%%%%%%%%%%%%%%
\subsection{Trustworthiness and Trust}
\label{sec:trustworthinessTrust}
When looking at literature related to (X)AI, \textbf{it is important to make a clear distinction between trust and trustworthiness}. 
Trust can be defined as an attitude that a stakeholder has towards a system \cite{Kastner.2021}, while trustworthiness is a property of the system that justifies to trust the system \cite{Toreini2020}.
One complication in the current literature, is that these concepts are not always clearly defined.
In some cases trust -- the attitude of a user -- and trustworthiness -- the property of the system -- are not clearly differentiated, or rather used interchangeably.
For example, \citet{BarredoArrieta.2020} describe trustworthiness as the confidence that a model will act as intended when facing a given problem, which is a fitting description of trust.
The differentiation is critical, because there are further factors apart from the system’s trustworthiness that also influence trust \cite{Toreini2020}. 

According to research on trust between humans (so-called trustor and trustee), trustworthiness is characterized by the trustee's ability, i.e. competence or expertise in the relevant context, the trustee's benevolence towards the trustor, and the trustee's integrity towards principles that the trustor finds acceptable \cite{Mayer.1995}.
Furthermore, a high level of these three factors of trustworthiness does not necessarily lead to trust, and trust can also occur in situations where lesser degrees of trustworthiness are present \cite{Mayer.1995}. 

For example, a meta-analysis \cite{kaplan2023trust} identified the expertise and personality traits of a person interacting with AI as significant predictors for trust in AI. 
Moreover, cultural differences, such as individualism and power relations within a culture, can influence a person's propensity to trust automation \cite{Chien2016, Chien2015}. Other factors influencing trust in automation may be the type of technology (e.g. using a DNN (deep neural network) blackbox model \cite{Bansal.2021} or decision tree \cite{Zhang2020}), the complexity of the task (e.g. tasks of varying cognitive load \cite{Wang2022, Jiang2022}), and perceived risk (e.g. using AI in a medical application \cite{Bussone2015}) \cite{Schaefer2016}. 

For these reasons, trust is influenced, but not determined by the trustworthiness of the system. Even the most trustworthy model will not be trusted in every case by every person. Vice versa, persons may -- and often do -- trust an untrustworthy model.

%%%%%%%%%%                          %%%%%%%%%%%%%%%
\subsection{Disuse, overtrust, and reliance in AI} 
\label{sec:reliance}

In the previous sections, we have discussed the use of trustworthy AI and explainability methods in preventing both disuse of AI systems as well as users overtrusting a useful yet imperfect AI.
Additionally, we have looked at the apparent connection between the objective of trust in AI, trustworthy AI, and explainability methods. 
In order to make a connection between trust and the concerns of disuse and overtrust, in this part we draw a connection between concerns of trust in AI and other (earlier) forms of automation \cite{lee2004trust, Hoff.2015}. 
While some issues and concerns may be specific to the AI context \cite{Glikson.2020}, the general concerns surrounding user reliance in technology have been a recurring theme over the last decades \cite{Kohn.2021}.
In the following, both work on trust in AI and trust in automation is considered. 
AI systems can be regarded as a subcategory of automation, and overarchingly we will refer to both as trust in a(n automated) system.

In the field of trust in automation the prevention of disuse and overtrust has been targeted by \textbf{ensuring appropriate trust} or calibrated trust.
\citet{McBride2010} as well as \citet{McGuirl2006} define appropriate trust as the alignment between perceived and actual performance of an automated system. This relates to a user's ability to recognize when the system is correct or incorrect and adjust their reliance on it accordingly. 
Within their model for trust calibration in human-robot teams, \citet{Visser.2020} define calibrated trust as given when a team member's perception of trustworthiness of another team member matches the actual trustworthiness of that team member.
If this is not given, either `undertrust', which leads to disuse, or `overtrust' can occur \cite{Visser.2020, Parasuraman1997}.

The aim of trust calibration by \citet{Visser.2020} is to assure a healthy level of trust and to avoid unhealthy trust relationships. 
Thereto, they establish a process of trust calibration which accompanies collaboration by establishing and continuously re-calibrating trust between the team members. To prevent people from overtrusting, so-called, trust dampening methods are to be applied. 
According to the authors, these methods are especially worthwhile in interactions with machines and robots, as humans have a tendency to expect too much from automation \cite{Visser.2020}. The authors recommend to present exemplary failures, performance history, likelihood alarms, or provide information about the system's limitations. Moreover, they make the connection to the expanding field of XAI arguing that explanation activities can help with calibrating trust. 

\textbf{Reliance in the AI context} can be understood as a human decision or action that considers the decision or recommendation of an AI. Trust is an attitude that benefits the decision to rely, as it has a critical role for human reliance on automation \cite{Hoff.2015}. So, beneath the desideratum of increased appropriate trust lies the desideratum of increased appropriate reliance. To rely appropriately, one would consider correct decisions or recommendations of an AI and would disregard false ones. 
Trust does not lead to this because trust is not only influenced by the correctness, i.e. the performance of an AI. 
According to \citet{Hoff.2015}, the performance of an automated system is similar to the trustee's ability in interpersonal trust, and the process and purpose of an automated system are analogous to benevolence and integrity. On top of that and as mentioned before, trust is not fully determined by trustworthiness.

\begin{figure}[h!]
    \centering
   % \fontsize{14}{17}\selectfont
    \includegraphics[width=\textwidth]{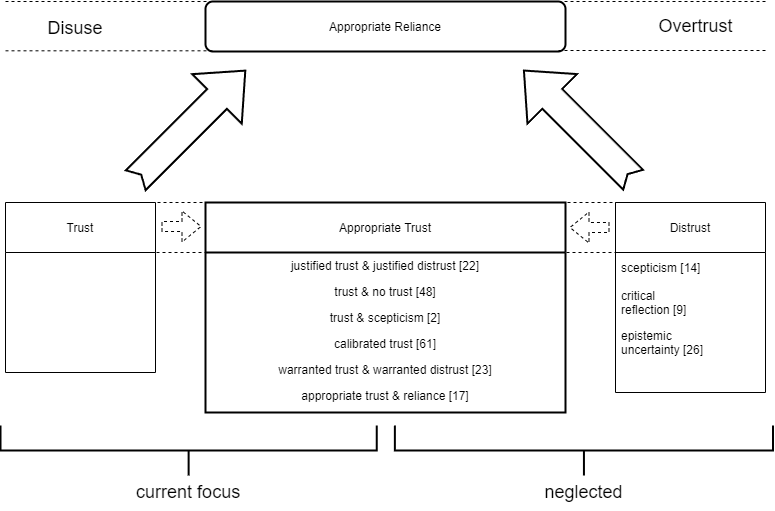}
    %\resizebox{\linewidth}{!}{\includesvg[]{approprReliance_version1.2-citations.svg}}
    \caption{Underlying desideratum of appropriate trust in AI and its relation to appropriate reliance.}
    \label{fig:approprTrust-Reliance}
\end{figure}

In other words, current improvements in automated systems, like XAI methods, are regarded as beneficial for \textbf{appropriate reliance} by preventing disuse and overtrust. Ideally, appropriate reliance should be achieved by fostering appropriate trust. Implied by the appropriateness of trust is that neither blind trust leading to overtrust, nor blind distrust leading to disuse is wanted. 
Several phrasing of this underlying notion of appropriate trust can be observed across the literature, which often entail trust and terms that can be summarized under distrust (see Fig. \ref{fig:approprTrust-Reliance}).

\nocite{Gaube.2021}
Yet, how can such an appropriate trust, its influencing factors, and the relation to appropriate reliance be conceptualized?
More recent trust research highlights a distinction that might be of interest here. Several researchers provided evidence that trust and distrust are two related, yet separate dimensions \cite{Lewicki.1998, Vaske.2016, HarrisonMcKnight.2001, Benamati.2006}.
We assume that this separation of trust and distrust might help solving the conceptual issue and propose that \textbf{we need an understanding of appropriate trust and healthy distrust}. The psychological underpinnings of our hypothesis will be detailed in the next sections.

%%%%%%%%%%%%%%%%%%%%%%%%%%%%%%%%%%%%%%%%%%%%%%%%%%%%%%%%%%%
%%%%%%%%%%                                 %%%%%%%%%%%%%%%%
\section{What is trust?} 
\label{sec:Trust}

Early, influential work on trust by sociologist \citet{luhmann2009a} defines trust as a mechanism for reducing the complexity of social interactions. Within social interactions, multiple goals and motives can be present, and multiple interpretation are possible with varying truth. According to the author, to decide which interpretation to follow and how to act upon them, this complexity needs to be reduced. By trusting, a person engages in the interaction as if only certain interpretations are possible (e.g., taking things at face value), and thus rendering the interaction less complex \cite{luhmann2009a}. 
In analogy, trust is also important in human-AI interactions, because of the involved risk caused by the complexity and non-determinism of AI \cite{Glikson.2020}. Similarly, \citet{Hoff.2015} argue that trust is not only important to interpersonal relations but can also be defining for the way people interact with technology. 

The typical conceptualisation of trust within (X)AI research regards trust as one end of a single dimension with distrust being the opposing end. The Integrative Model of Organizational Trust by \citet{Mayer.1995}, which elaborates this conceptualisation, is a prominent basis for trust in AI and automation \cite{Stanton.2021}.
\citet{Mayer.1995} define trust as ``[...] the willingness of a party to be vulnerable to the actions of another party based on the expectation that the other will perform a particular action important to the trustor, irrespective of the ability to monitor or control that other party" \cite{Mayer.1995}. Based on this definition, they differentiate between factors that contribute to trust, trust itself, the role of risk, and the outcomes of trust.

The extent of a person's willingness to trust is influenced by the trustor's propensity to trust and the trustor's perception of the trustworthiness of the trustee. 
To separate trust from related constructs \citet{Mayer.1995} highlight the importance of risk and vulnerability for trust. They argue that, if a situation does not involve a form of vulnerability to the trustor, cooperation can occur without trust. Similarly, if a trustor does not recognize and assume any risk, the trustor is in a situation of confidence and not of trust. 
Thus, trust serves the purpose of reducing complexity in an interaction, and for trust to be present, a form of vulnerability and risk is required. 

Trust in a system and trust defined by \citet{Mayer.1995} share that they influence the willingness to rely and the situational requirements of risk and vulnerability for them to be of importance. 
According to \citet{Muir1996} users used automated systems they trust but not those they do not trust. \citet{Lee1994} state that operators did not use automation systems if their trust in them was less than their own self-confidence. 

Drawing from \citet{Mayer.1995}, definitions of trust in automation also consider the necessity of uncertainty (i.e., risk) \cite{Hoff.2015, lee2004trust} and vulnerability \cite{lee2004trust, Kohn.2021}. Trust in automated systems ``plays a leading role in determining the willingness of humans to rely on automated systems in situations characterized by uncertainty" \cite{Hoff.2015}, and is defined as ``[...] the attitude that an agent will help achieve an individual’s goals in a situation characterized by uncertainty and vulnerability" \cite{lee2004trust}. 

Returning to the desired appropriate reliance via appropriate trust (Section \ref{sec:reliance}) and combining it with the discussed insights on trust in automated systems, which drew from \citet{Mayer.1995}, we see a problem. 
To reiterate, the desideratum is to prevent either the case in which a person relies on the AI, even though it was wrong (overtrust), or the case, where a person does not rely on the AI, even though it was correct (disuse).  
Fostering trust, i.e., increasing the willingness to rely, mitigates the problem of disuse. However, for mitigating overtrust, not an absence of the willingness to rely, but the ability to identify reasons not to rely is needed. Trust in \citeauthor{Mayer.1995}'s model does not entail this, as they define trust “[…] irrespective of the ability to monitor or control that other party”. 
\citeauthor{Mayer.1995}'s influential work on trust demonstrates the difference between trust and trustworthiness, but for the mitigation of overtrust, their model does not provide a basis to proceed. For our proposition to resolve this, not only trust but also distrust is needed.

%%%%%%%%%%%%%%%%%%%%%%%%%%%%%%%%%%%%%%%%%%%%%%%%%%%%%%%%%%%
%%%%%%%%%%                                 %%%%%%%%%%%%%%%%
\section{The importance of distrust} 

Distrust is often connotated negatively \cite{Lewicki.1998, Vaske.2016} and sometimes explicitly considered something to be avoided \cite{Frison.2019, Seckler.2015}, or at least implied to be avoided when focusing the sole strengthening of trust.
Yet considering the imperfection of contemporary ML models, distrust towards erroneous predictions and towards explanations that indicate them is not to be avoided, but fostered. 
Otherwise, a neglect of distrust remains, which is serious because it renders potential positive consequences of distrust invisible.

In a study by \citet{McKnight.2004} the disposition to distrust predicted high-risk perceptions better than the disposition to trust did. For their study context of online expert advice sites they suggest that future research should study dispositional trust and also dispositional distrust. 
Psychological studies also point to the benefit of considering distrust by identifying positive consequences of distrust.
Distrust or suspicion led, for example, to an increase of creativity \cite{Mayer.2011} or a reduction of the correspondence bias \cite{Fein.1996}. 
Moreover, a series of studies by \citet{Posten.2021} showed an increase of memory performance in their distrust condition as opposed to a trust or control condition. 
\citet{Vaske.2016} identified a potential of distrust to improve critical reflection and innovation in the context of working in an organisational setting. 

Looking at potential underlying mechanisms of distrust, \citeauthor{Mayo.2015}'s \cite{Mayo.2015} review introduces a so-called distrust mindset as an explanation for the positive effects of distrust. The distrust mindset leads to an activation incongruent and alternative associations, which aligns very well with the increase of creativity, reflection, and innovation. 
According to \citet{Posten.2021}, trust triggers a perception focus on similarities that makes it harder to remember single entities. Distrust shifts the perception focus towards differences and, therefore, increases memory performance. 
Interestingly, in one of their studies \citet{Posten.2021} observe a higher acceptance of misinformation in a trust condition, underlining the potential problem of the current trust focus in the (X)AI context and the danger of overtrust. 

A conceptual example of how trust and distrust can be targeted is provided by \citet{Hoffman.2018} in their work on measuring trust in XAI. 
They advocate that people experience a mixture of justified and unjustified trust, as well as justified and unjustified mistrust. Ideally, the user would develop the ability to trust the machine in certain task, goals, and problems, and also to appropriately mistrust\footnote{Mistrust is used as a synonym for distrust in this paper.} the machine in other tasks, goals, and problems. 
This ideal scenario requires them to be able to decide when to trust or to correctly distrust, when scepticism is warranted.
In sum, although often connotated negatively, distrust also has positive consequences and its own merits separate from those of trust.

%%%%%%%%%%%%%%%%%%%%%%%%%%%%%%%%%%%%%%%%%%%%%%%%%%%%%%%%%%%
%%%%%%%%%%                                 %%%%%%%%%%%%%%%%
\section{Distrust as a separate dimension}
While distrust is often regarded as the opposite of trust, the concept of a one-dimensional view of trust and distrust is being questioned and not widely accepted \cite{Schweer.2009, Lewicki.1998, Guo.2017, Schoorman.2007}.  
In the two-dimensional approach, by definition, low trust is not the same as high distrust, and low distrust is not equal to high trust \cite{Lewicki.1998}. This allows the coexistence of trust and distrust. 
Among others, trust is characterized by hope, faith, or assurance, and distrust by scepticism, fear, suspicion or vigilance \cite{Benamati.2006, Cho.2006, Lewicki.1998}. \citet{Lewicki.1998} exemplify the separation of trust and distrust by contrasting low trust with high distrust. The authors regard expectations of beneficial actions being absent or present as antecedent to trust, and expectations of harmful actions being absent or present as antecedent to distrust. 
If the former is absent, low trust is expressed by a lack of hope, faith, and confidence. If the later is present, high distrust is expressed by vigilance, skepticism, and wariness. The combination of high trust and high distrust is described by the authors with a relationship in which opportunities are pursued while risks and vulnerabilities are continually monitored. 

When reviewing research that draws from two-dimensional approaches, concepts and terms like critical trust, trust but verify, and healthy distrust are used \cite{Poortinga.2003, Lewicki.1998, Vaske.2016}.
These align well with the problem of mitigating overtrust, yet little consideration of the two-dimensional view on trust and distrust can be found when trust is considered in the technology context.

One, at least partial, reason for this is found in the field of organisational psychology. \citet{Vaske.2016} discusses the trajectory of the conceptual debate about trust and distrust within the organizational context. He describes that most of the earlier work on trust falls into the category of the one-dimensional approaches. From the mid 80s onward, doubt increased towards the one-dimensional approach, which was considered too simplistic \cite{Vaske.2016}. 
Yet, efforts to resolve this debate and empirically test it remain scarce, while work on the two-dimensional approach mostly reproduces common-sense assumptions instead of providing empirical evidence \cite{Vaske.2016}.

Furthermore, \citet{Vaske.2016} points out that only the concept trust has a good theoretical background and is well researched.
Distrust remains in the state of conceptual debate and is given little research attention. As a consequence, even in the field of organizational psychology, in which the conceptual critique towards the one-dimensional approach is the most visible, applied work still relies mostly on the model by \citeauthor{Mayer.1995} \cite{Vaske.2016}.
%%% TiA & (X)AI also relies on Mayer1995 model, resulting in similar neglect of distrust
As highly influential work on trust in automation \cite{lee2004trust, Hoff.2015} also draws from \citet{Mayer.1995}'s model, which then was taken as a starting point in the context of trust in XAI \cite{Thiebes.2021}, these fields inherited the focus on trust and the neglect of distrust.

Regardless of evidence for the two-dimensional conceptualisation, uni-dimen\-sional scales are the common form to evaluate trust in automation \cite{Kohn.2021}.
Of those, the Checklist for Trust between People and Automation \cite{Jian.2000} is the most frequently used one \cite{Kohn.2021}. 
This checklist measures trust and distrust as polar opposites along a single dimension. Five of the 12 items (statements rated by the user) measure distrust. In practice, these items are often reverse-scored and summed with the trust items to form one trust score, which was also suggested by the original authors of the scale \cite{Spain.2008}. A critical validation attempt of this scale by \citet{Spain.2008} compared a one-factor model (indicating the polar opposites along a single dimension) and a two-factor model. 
This factor analysis provided evidence for the conceptualization of trust and distrust as separate, yet related constructs \cite{Spain.2008}. Thus, reverse scoring distrust items to then form a sum score with the trust items entails a problematic entanglement of the two factors identified by \citet{Spain.2008} and disregards the incremental insight by measuring trust and distrust individually.

The merit of considering trust and distrust as separate dimension has been identified across different sub-fields of human-technology interaction \cite{Kohn.2021, HarrisonMcKnight.2001, McKnight.2004, Benamati.2006, Ou.2010, Fang.2015, Thielsch.2018}.
A difference between dispositional trust and dispositional distrust was observed in the context of online expert advice \cite{McKnight.2004}, and trust and distrust co-existed as distinct construct in the context of online banking \cite{Benamati.2006} and online shopping \cite{Ou.2010}. A study on website design showed that trust and distrust are affected by different antecedents, and the performance of a trust-aware recommender system was improved by not only predicting trust but also distrust \cite{Fang.2015}. \citet{Thielsch.2018} investigated work-related information systems and also identified trust and distrust as related yet separate influences on different outcome variables. 

Some authors \cite{Fang.2015, Ou.2010, Thielsch.2018} emphasize that they are, to the best of their knowledge, the first to consider not only trust but also distrust in their field. Additionally, with a distance of two decades \citet{HarrisonMcKnight.2001} and \citet{Kohn.2021} both argue in favour for considering trust and distrust in the technology. 
This indicates a lack of generalization on the conceptualisation of trust and distrust in context of technology. The impression arises that within individual sub-fields at different points in time, the potential merit of considering trust and distrust is identified, and only some first steps are taken towards it. 
Some of the studies that take these first steps still only partially separate trust and distrust, which may hinder generalization.

For instance, \citet{Benamati.2006, Fang.2015} make the distinction between trust and distrust only on a superficial level because they relate trust and distrust to the same antecedents \cite{Benamati.2006, Fang.2015} and consequences \cite{Benamati.2006}. Therefore, they do not fully acknowledge \citeauthor{Lewicki.1998}'s proposition \cite{Lewicki.1998} of different antecedents and consequences of trust and distrust. 
In both cases, the authors themselves suggest addressing these limitations in future research. \citet{Fang.2015} suggest to predict trust and distrust from different antecedents. \citet{Benamati.2006} entertain the degree of monitoring as plausible outcome for distrust instead of the intention to use, which they had used as an outcome for both trust and distrust.

%%%%%%%%%%%%%%%%%%%%%%%%%%%%%%%%%%%%%%%%%%%%%%%%%%%%%%%%
%%%%%%%%%%                              %%%%%%%%%%%%%%%%
\section{Conclusion} 
\label{sec:needHealthyDistrust}

To summarize, a focus on trust and the neglect of distrust is evident in research about trust in (X)AI, trust in automation, and in trust research in organizational psychology. 
Some examples of considering both trust and distrust can be identified in different sub-fields about interaction with technology.
The underlying idea of these studies, that trust does not suffice, is strengthened by the examples of positive consequences of distrust. 

The notion of appropriate trust in current (X)AI research also acknowledges that trust alone does not suffice.
However, by aiming for appropriate trust, a crucial ambiguity remains, because stating that it is not appropriate to trust allows for two interpretations: either that one does not trust, or that one distrusts.
The same distinction has to be made when returning to the problems of disuse and overtrust. By increasing trust, the problem of disuse can be mitigated, as the willingness to rely increases. While it may be true that a lower willingness to rely, a lower trust, would decrease the likelihood of overtrust, there would only be less reliance overall. 
To this point, to mitigate overtrust, reliance should be prevented if, and only if, it would be wrong to rely. To conceptualize this distinction we regard distrust as relevant. 
Thus, \textbf{we propose the consideration and evaluation of both trust and distrust to achieve appropriate reliance in AI by mitigating both disuse and overtrust}.
We consider our proposition to be in line with the underlying motivation of appropriate trust (see Fig. \ref{fig:approprTrust-Reliance}). 

However, the term and work on it is often too keen on trust and disregards the critical review of distrust. 
The aim of our proposition is to ensure trust plus a healthy amount of distrust. 
Being inherently imperfect, contemporary AI will benefit from healthy distrust, as this entails a more conscientious usage of AI. 
With healthy distrust, a user would have a warranted critical stance towards the AI's outputs without being outright distrustful towards using the AI at all.

For instance, current AI chatbots generate plausible sounding and highly convincing texts.
A situation in which overtrust is arguably a more prominent issue than disuse. 
Developing or implementing methods to support the identification of possibly wrong outputs would be a sensible approach to mitigate overtrust. 
Ideally, such methods would foster the user's wariness towards and the user's monitoring of the system, i.e. characteristics of distrust. 
If these envisioned methods were to only be evaluated by measuring trust, different results would be plausible. On the one hand, a decrease in trust could be observed. On the other hand, observing no change in trust or even an increase in trust, despite errors being identified more easily, would also be plausible. 
Firstly, the trust measurement may not change because only characteristics of distrust would be affected by the methods sketched above.
Secondly, trust may increase if the user considered the output as correct regardless of the supported ability to monitor. This should occur by giving the user the opportunity to distrust. The user would have better ability to examine the output and could verify his trust, resulting in a stronger willingness to rely. Such potential impacts of distrust will not be directly noticed when only measuring trust. 
In these hypothetical scenarios additionally evaluating distrust would provide more insights into the user's attitude when interacting with AI.

Generally, evaluating both trust and distrust could help to clarify the mixed and inconclusive results in empirical research on the explainability-trust hypothesis (Section \ref{sec:explainability-trust}). As explanations have the two utilities of identifying both correct as well as wrong outputs, explanations may influence both the user's trust and distrust.
To conclude, in the aim of appropriate reliance on trustworthy (X)AI both trust and distrust should be considered.

%%%%%%%%%%%%%%%%%%%%%%%%%%%%%%%%%%%%%%%%%%%%%%%%%%%%%%%%
%%%%%%%%%%                              %%%%%%%%%%%%%%%%
\section{Open questions and future work}

The conceptualisation of distrust in AI and the empirical identification of its antecedents and consequences still needs further empirical research.  
Aspects that constitute healthy distrust have to be identified as well.
To progress this, future work should not only separate trust and distrust on a superficial level, but also investigate whether and which individual antecedents and consequences of trust and distrust are relevant. 
Studies on trust and distrust in the (X)AI context need to continue to draw from the established work on (dis)trust without intermixing it with common-sense reasoning on trust and distrust.

Furthermore, for researchers to be able to appropriately compare what effects different (X)AI methods, design, and systems have on user trust and distrust standardized ways of measuring them as separate dimensions have to be created and validated. 
With overcoming such conceptual and methodological issues, the two-dimensional concept of trust and distrust can be validated more convincingly. Thereby the lack of generalization can be addressed and an improved starting point for further research on both trust and distrust can be established. 

%%%%%%%%%%%%%%%%%

\bibliographystyle{splncs04nat}

\bibliography{bibliography.bib}

\end{document}